\documentclass[10pt]{article}
\usepackage{cite}
\usepackage{mathptmx}       
\usepackage{helvet}         
\usepackage{courier}        
\usepackage{type1cm}        
\usepackage{makeidx}         
\usepackage{graphicx}        
\usepackage{multicol}        
\usepackage[bottom]{footmisc}
\makeindex             

\begin{document}

\title{Predicting evolution and visualizing high-dimensional fitness landscapes}
\author{Bj\o rn \O stman$^{1,2,\star}$ and Christoph Adami$^{1,2,3}$ \\ \\
$^1$Department of Microbiology and Molecular Genetics\\
$^2$BEACON Center for the Study of Evolution in Action\\
$^3$Department of Physics and Astronomy\\ \\
Michigan State University, East Lansing, MI\\
$\ast$ E-mail: ostman@msu.edu}
%
%
\maketitle

\abstract{The tempo and mode of an adaptive process is strongly determined by the structure of the fitness landscape that underlies it. In order to be able to predict evolutionary outcomes (even on the short term), we must know more about the nature of realistic fitness landscapes than we do today. For example, in order to know whether evolution is predominantly taking paths that move upwards in fitness and along neutral ridges, or else entails a significant number of valley crossings, we need to be able to {\em visualize} these landscapes: we must determine whether there are peaks in the landscape, where these peaks are located with respect to one another, and whether evolutionary paths can connect them. This is a difficult task because genetic fitness landscapes (as opposed to those based on traits) are high-dimensional, and tools for visualizing such landscapes are lacking. In this contribution, we focus on the predictability of evolution on rugged genetic fitness landscapes, and determine that peaks in such landscapes are highly clustered: high peaks are predominantly close to other high peaks. As a consequence, the valleys separating such peaks are shallow and narrow, such that evolutionary trajectories towards the highest peak in the landscape can be achieved via a series of valley crossings.}

\section{Usefulness of the metaphor }
\label{sec:1}
The structure of realistic fitness landscapes continues to be an active area of research. When discussing the ÒmetaphorÓ of the fitness landscape, researchers typically mean the one- or two-dimensional representation of fitness as a function of either genotype or phenotype. This is a powerful image that immediately evokes how an evolving population may change, given population size and mutation rate: the familiar image of a landscape with hills and valleys initially described by Wright~\cite{Wright1982,Wright1932,Whitlock1997,Frankeetal2011,KauffmanLevin1987,Jain2007}
 affects our intuition about evolutionary dynamics. However, as genes or traits do not affect fitness in isolation, considering the fitness contribution of only one or two genes may focus on too small a part of the fitness landscape to make relevant predictions about how populations actually evolve. Instead, it is possible that many more dimensions are required to understand adaptation, which makes a straightforward visualization of the landscape impossible. Is the fitness landscape metaphor still instructive if many genes contribute to fitness, that is, if the fitness landscape is high-dimensional?   This concern has been raised multiple times~\cite{Whitlocketal1995,Kaplan2008,Pigliucci2008}. In particular, it has been questioned whether the peaks separated by valleys commonly seen in one- and two-dimensional fitness landscapes exist at all in higher dimensions. An argument by Gavrilets~\cite{Gavrilets1997,Gavrilets2004}  states that when the number of dimensions is very high, peaks disappear because they become connected by neutral ridges. This mathematical argument was presented as a solution to the problem of valley-crossing~\cite{Frankeetal2011,Jain2007,KauffmanLevin1987,Whitlock1997,Wright1982}. Indeed, the discussion about the structure of fitness landscapes is intimately linked to the dynamics of adaptation. After all, WrightÕs shifting balance theory~\cite{Wright1931,Wright1932}---which suggest that peak shifts occur via random drift during times when population sizes are small---is only relevant if valleys actually need to be crossed. Fisher, on the other hand, figured that landscapes change often enough that populations can simply wait on a local peak until beneficial mutations became possible~\cite{Fisher1930}, after which the new peak will be climbed via a series of slightly beneficial mutations. If, on the other hand, all high-fitness genotypes are actually connected by ridges of genotypes with equal fitness, populations could then readily evolve along these ridges, and thus the problem of having to cope with deleterious mutations disappears. 

However, two important arguments exist against this solution. First, even if these ridges exist, it is not clear whether they are numerous enough that they are preferred by an evolving population. As these ridges appear only once the fitness landscape is high-dimensional, the number of possible paths for the population to take is exponentially large. Only a very small fraction of these may be neutral ridges~\cite{Gavrilets2004}, and when the mutation supply is sufficient, it is more likely that the population will traverse a short (i.e., single-step) as well as shallow valley, rather than taking a longer neutral path (see results below). Secondly, when all high-fitness peaks are connected by neutral ridges as in Gavrilets$Õ$s Holey Landscape model, the high-fitness genotypes must permeate all areas of genotype space, and as a consequence are not clustered in genotype space~\cite{Gavrilets2004}. However, it is not at all clear whether we should expect such a homogenous distribution of high fitness peaks. Indeed, just as Kauffman originally hypothesized~\cite{Kauffman1993}, we have recently shown that---at least in the NK model of epistatic fitness landscapes---peaks are not evenly distributed, but that they instead cluster so that high fitness peaks are closer to each other than they are to other low fitness peaks~\cite{Ostmanetal2010}. In that work we studied sequences (haplotypes) with $N=20$ loci only, which does not meet Gavrilets's criterion for a high-dimensional space. However, it is possible to look for the same phenomenon in landscapes with a much higher dimensionality, such as the landscape created by digital evolution environments such as Avida~\cite{Adami1998,OfriaWilke2004,Adami2006}. 

As for empirical landscapes, Whitlock et al.~\cite{Whitlocketal1995} asserted that it may not be possible to conclude beyond a shadow of a doubt that complex biological fitness landscapes are epistatic and contain multiple peaks, but that the evidence is very strong that they do. Since that review, many more empirical landscapes have been reconstructed, and the evidence for multiple peaks, strong epistatic interactions between genes, and local optima continues to increase~\cite{Lunzeretal2005,Beerenwinkeletal2007,ElenaLenski1997,Hinkleyetal2011,PittFerre-DAmare2010,Kouyosetal2012,Kryazhimskiyetal2011,KvitekSherlock2011}.
If biological fitness landscapes contain multiple peaks, the problem of just how populations manage to cross fitness valleys is real. Besides the now standard solutions of Fisher and of Wright, we now know that a high mutation-supply rate (product of population size and mutation rate) enables populations to cross valleys via compensatory mutations, despite the transient reduction in fitness~\cite{Weissmanetal2009,Ostmanetal2012}. The more mutations the population incurs, the higher the level of standing genetic variation, and the higher the chance that inferior genotypes will give rise to offspring that are lucky enough to move closer to an adjacent peak. Of course, if the mutation-supply rate is low, then deleterious mutations are rarely tolerated, and in the extreme case of mutations being so rare that each mutation goes to fixation before the next is available, valleys of even moderate fitness loss cannot be crossed. But many biological systems have mutation-supply rates high enough that there are always plenty of opportunities for organisms of lower fitness to have offspring with mutations. In fact, some populations behave like quasispecies~\cite{Eigenetal1988,Eigen1971}: typical examples are viruses~\cite{BurchChao2000} and even bacteria~\cite{CovacciRappuoli1998}. Such populations are characterized by extended clouds of genotypes containing many genetic variants, and these can easily cross valleys as the fixation probability of a deleterious mutant connected to the quasispecies is nonvanishing~\cite{Wilke2003}. 
No matter how valleys are crossed during adaptation, we can now be almost certain that fitness landscapes contain distinct peaks, even in the multidimensional space of biological organisms. Therefore, we contend that the fitness landscape metaphor, exemplified by the image of hills and valleys in two dimensions, is useful in guiding our intuitions about evolutionary dynamics. While a high-dimensional landscape may differ quantitatively from the familiar picture of rolling hills and valleys, we believe that the idea that there are peaks and valleys in a biological fitness landscape is unlikely to be refuted.

\section{Visualizing fitness landscapes}

Even though low-dimensional fitness landscapes can sometimes serve as an appropriate visual for landscapes of higher dimensions, the use of other methods for visualizing multidimensional fitness landscapes is currently a topic of much interest. Techniques for visualization include different methods for creating two-dimensional representations of high-dimensional space~\cite{WilesTonkes2006,McCandlish2011}. These methods are applied when complete knowledge of the fitness of all genotypes exists. Due to the inherent multi-dimensionality of genotypes, obtaining full information for biological systems is not feasible. The fitness landscapes of limited areas of genotype space have recently been constructed for {\it E. coli}~\cite{Chouetal2011,Khanetal2011}, RNA ~\cite{PittFerre-DAmare2010},  HIV-1~\cite{Martins2012}, and {\it Aspergillus niger}~\cite{Frankeetal2011}. 
	This interest in reconstructing and visualizing fitness landscapes stems from the promise that knowledge of this function (i.e., fitness) holds, namely that in combination with population size and mutation rate, evolutionary dynamics becomes predictable. But what exactly does this predictability achieve? Which evolutionary events can we predict? Given the stochasticity inherent in evolutionary dynamics, does knowledge of the three core evolutionary parameters---population size, mutation rate, and fitness landscape---actually lead to testable predictions? Do we currently know enough about fitness landscapes that they can be used to make predictions about the likely future paths?
	
	The most obvious benefit from knowing the structure of a fitness landscape is a better understanding of adaptation. For example, just knowing the topology of the local fitness landscape is enough to tell whether the population will experience adaptation or not. If the population is already located on a fitness peak, then it will remain there provided that the local structure is such that mutation-supply rate is too low for adjacent peaks to be reached. In the event that there are adjacent peaks of higher fitness, crossing the valleys in between would result in an adaptive relocation. As mentioned above, valley-crossing is only a hindrance to adaptation when the combination of low mutation-supply rate and deep and/or long valleys between peaks prohibits the population from taking advantage of the potential of deleterious mutations. 
	
	As speciation is a process intimately tied to evolutionary dynamics, it is also impacted by the structure of the fitness landscape~\cite{Gavrilets2004}. Genetic differences underlying multiple species necessarily result in those species occupying different areas of genotype space. If these areas are not separated by valleys, then intermediate genotypes (hybrids, in the case of sexually reproducing species) are of high fitness, and both initiation and maintenance of speciation will be less likely. A multi-peaked landscape therefore promotes speciation~\cite{DoebeliDieckmann2000}. Given two adjacent fitness peaks, a population in the vicinity of both can adapt to either optima. If there are no physical or genetic barriers to competition within the population, the population will end up on one or the other (most likely the highest of the two, but not necessarily so). However, if such barriers to competition do exist, the population may permanently split into two, eventually leading to ecological speciation~\cite{NosilHarmon2009}. If the peaks correspond to genomic regions or traits that directly affect reproduction, speciation will be more likely to occur. Additionally, the more rugged the fitness landscape is with respect to traits that affect resource use and reproductive isolation, the higher the chance that evolutionary branching will lead to speciation. Consequently, the more dimensions relevant to fitness that are under scrutiny, the more precisely we can predict the evolutionary outcome with respect to speciation.
	Both adaptation and speciation are thus affected by how rugged the fitness landscape is. The more epistatic interactions between loci, the more rugged the landscape is, and the more peaks there are. Not only does the number of peaks matter, spatial correlations between peak genotypes could also affect adaptation and speciation. If peaks are not evenly distributed in genotype space, but are clustered in some fashion, close proximity of peaks can facilitate peak shifts and thereby increase the likelihood of adaptive shifts and ecological speciation. For sexually reproducing species, speciation may require that valleys in between peaks are deep and wide enough to cause hybrid sterility or breakdown~\cite{Turellietal2001}. As we shall see below, there are reasons to believe that peak clustering is a generic feature of multi-peaked, epistatic fitness landscapes.

 \section{Landscape structure and peaks}

Studying the large-scale structure of biological fitness landscapes is difficult due to several factors. Measuring fitness is not trivial and experimentally manipulating organisms to explore large areas of genotype space is an arduous task. However, some numerical models have fitness landscapes where all genotypes can be enumerated. The NK model is such a landscape~\cite{KauffmanLevin1987,Kauffman1993}, in which the amount of epistatic interactions can be adjusted and ruggedness thereby varied in a controlled manner. 
	Briefly, $N$ is the number of bi-allelic loci in the genome, and $K$ is the number of neighboring loci that each locus interacts with. Fitness is calculated as the average fitness contribution of each locus, which in turn is taken from a table of uniform random numbers. In a $K$=0 landscape loci do not interact, and as a consequence the landscape contains only a single peak. Adaptation in such a landscape is predictable, as there is no other outcome than the population ending up on the peak (or around it, in case of a high mutation-supply rate). In contrast, for $K=N-1$, all loci interact with all other, and the fitness of neighboring genotypes are completely uncorrelated, creating a maximally rugged landscape. Both of these extremes have been mathematically investigated, but neither have much bearing on real fitness landscapes, which are rarely non-epistatic, and never completely lacking in fitness correlations~\cite{KauffmanLevin1987,Frankeetal2011}. The intermediate range of $K$, which creates varying degrees of epistasis and ruggedness, is difficult to study analytically, but lends itself easily to computational methods~\cite{Ostmanetal2012}.
\begin{figure}[!tb]
\centering
\includegraphics[scale=.6]{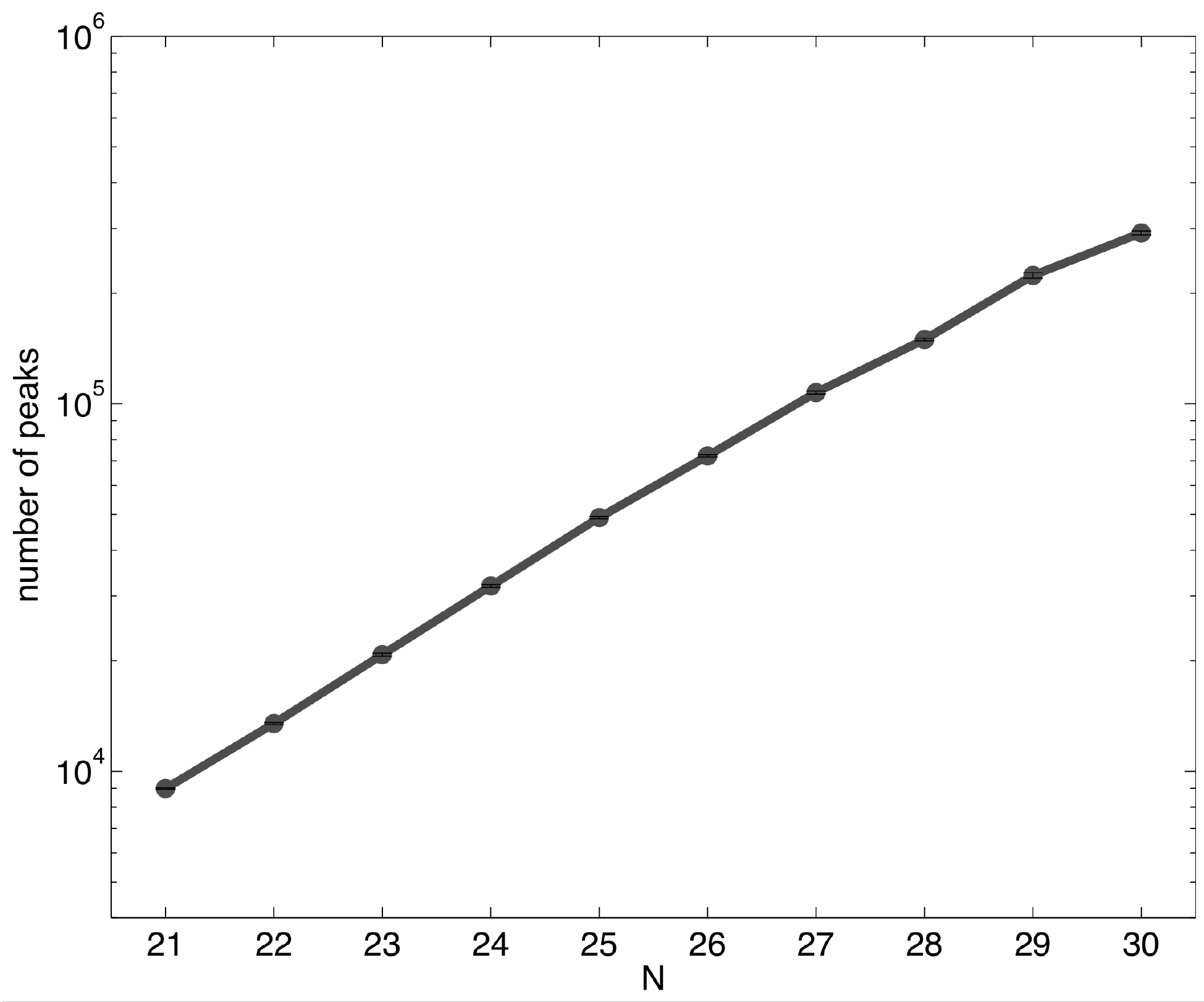}
\caption{In the NK landscape, the number of peaks increases exponentially as a function of the number of loci. Here shown for a highly rugged landscape, $K$=8, and $N$ between 21 and 30. Peaks are here defined as genotypes whose $N$ one-mutant neighbors all have lower fitness. Every datum is the average over five different landscapes, and the s.e.m. error bars in black are smaller than the markers. To the extent that it has been investigated, there is no indication that peaks cease to exist in multidimensional fitness landscapes, contra Gavrilets~\cite{Gavrilets1997,Gavrilets2004}.}
\label{fig:1}       
\end{figure}
	Here we present several landscape measures~\cite{Richter2008} that characterize the large-scale structure of the NK landscape, and whose specifics in turn affect adaptation and evolutionary dynamics. In accordance with the Massif Central hypothesis~\cite{Kauffman1993}, peaks are unevenly distributed in genotype space, and tend to form clusters. Several lines of evidence support this inference. First, defining peaks as genotypes whose one-mutant neighbors are all of lower fitness, we look at the dependence of the number of peaks on landscape dimensionality. As the dimensionality of the NK model increases with the number of loci, $N$, the number of peaks increases exponentially (Fig.~1). There is thus no indication that neutral ridges appear as the dimensionality of the NK model in increased. Looking at the spatial distribution of peaks (genotypes for which all $N$ mutational neighbors have a lower fitness), we found that peaks are located closer to each other than expected if they were distributed randomly~\cite{Ostmanetal2010}. The Hamming distance between peaks is lower than between a set of random genotypes (Fig.~2A), where the random control displaces peaks by randomly assigning them new genotypes without replacement. This correlation disappears for high $K$, but reappears when one only looks at the highest peaks (Figs.~2B-D). Moreover, the higher the peaks are, the more likely they are located near each other. This can be seen from the strong association between the fitness of a randomly selected peak and the peaks in its near vicinity. Looking at all peaks and comparing their fitness to the average fitness of peaks close by (Hamming distance of 2), a very strong correlation is evident (Fig.~3). The strength of the correlation depends on the ruggedness of the landscape, but even though it is weaker for $K$=4 than for $K$=2, it is evident from this analysis that peaks are more likely to be found near peaks of a comparable rather than much different height.
The observed effect of a strong correlation between fitness of the highest peaks (and the watering down of that correlation when the peaks of lower fitness are included in the analysis) suggests an interesting feature of multidimensional rugged landscapes. Peaks are more likely to be located near other peaks of similar fitness, so not only is there a Massif Central, but there also exist clusters of peaks across the NK genotype space. If we look only at peaks of fitness above a threshold $\theta$ (Fig.~2), the spatial correlation in genotype space is strong, but as we lower that threshold, more peaks appear and are spread out over more of genotype space. Additionally, as $\theta$ is lowered, more and more peaks become connected in a network where peaks are no further apart than a Hamming distance of two. We can observe a percolation phase transition for peak networks, where peak networks change from being disconnected to forming large networks that include nearly all peaks (Fig.~4). (Because a network of peaks connected by a Hamming distance of exactly two cannot be connected to peaks separated by a Hamming distance of three, there are two large networks in Fig.~4, which is why less than half or all peaks are found in the largest network.)
If such rugged landscapes contain peaks in large areas of genotype space, we may then ask just how much of genotype space the clusters of peaks inhabit. Are peaks distributed evenly such that all points in genotype space are equally likely to be near a peak, or are there large areas of genotype space that are devoid of peaks?
To test whether rugged landscapes contain peaks of low fitness distributed evenly in genotype space, we translated the bit-string NK genotypes into decimal numbers in the following way. Denote the binary string by the vector $\vec s$. The corresponding decimal genotype is then $d(\vec s)=\sum_i
s_i 2^{i-1}$, where the summation is over all indices of  $\vec s$. For example, the genotype 10010  is represented by the vector $\vec s=\{1,0,0,1,0\},$ and has the decimal genotype  $d(\vec s)=1\times2^0+0\times 2^1+0\times 2^2+1\times 2^3+0\times 2^4=9$. 

This formalism can easily be extended to non-binary strings, e.g., genotypes consisting of more ÒallelesÓ. For DNA, $d(\vec s)=\sum_i s_i4^{i-1}$ , so for example (with A=0, T=1, C=2, G=3, reading right to left) TATA corresponds to $d(\vec s)=0\times 4^0+1\times 4^1+0\times 4^2+1\times 4^4=68$, and the one-mutant GATA to $d(\vec s)=0\times 4^0+1\times 4^1+0\times 4^2+3\times 4^4=196$. As we can see from this example, neighboring genotypes with Hamming distance one are not necessarily neighbors in this decimal genotype space. A genotype with two mutations can be closer than a single-mutant, as with TCTT, which has $d(\vec s)=1\times 4^0+1\times 4^1+2\times 4^2+1\times 4^4=101$ . 
If there were no structure to how peaks are distributed in genotype space, the distribution of peaks in decimal genotype space is expected to appear even. However, despite the disconnect between neighboring genotypes in decimal space, plotting peak fitness as a function of the decimal genotype reveals an intriguing order not observed among random genotypes (Fig.~5). Peaks are irregularly distributed and large areas have no peaks. These basins of attraction~\cite{Ellison2000} clearly show that peaks are not spread evenly in genotype space, but cluster together in agreement with the analysis above (Figs.~2 and 3). 
%
\begin{figure}[!tb]
\centering
\includegraphics[scale=.6]{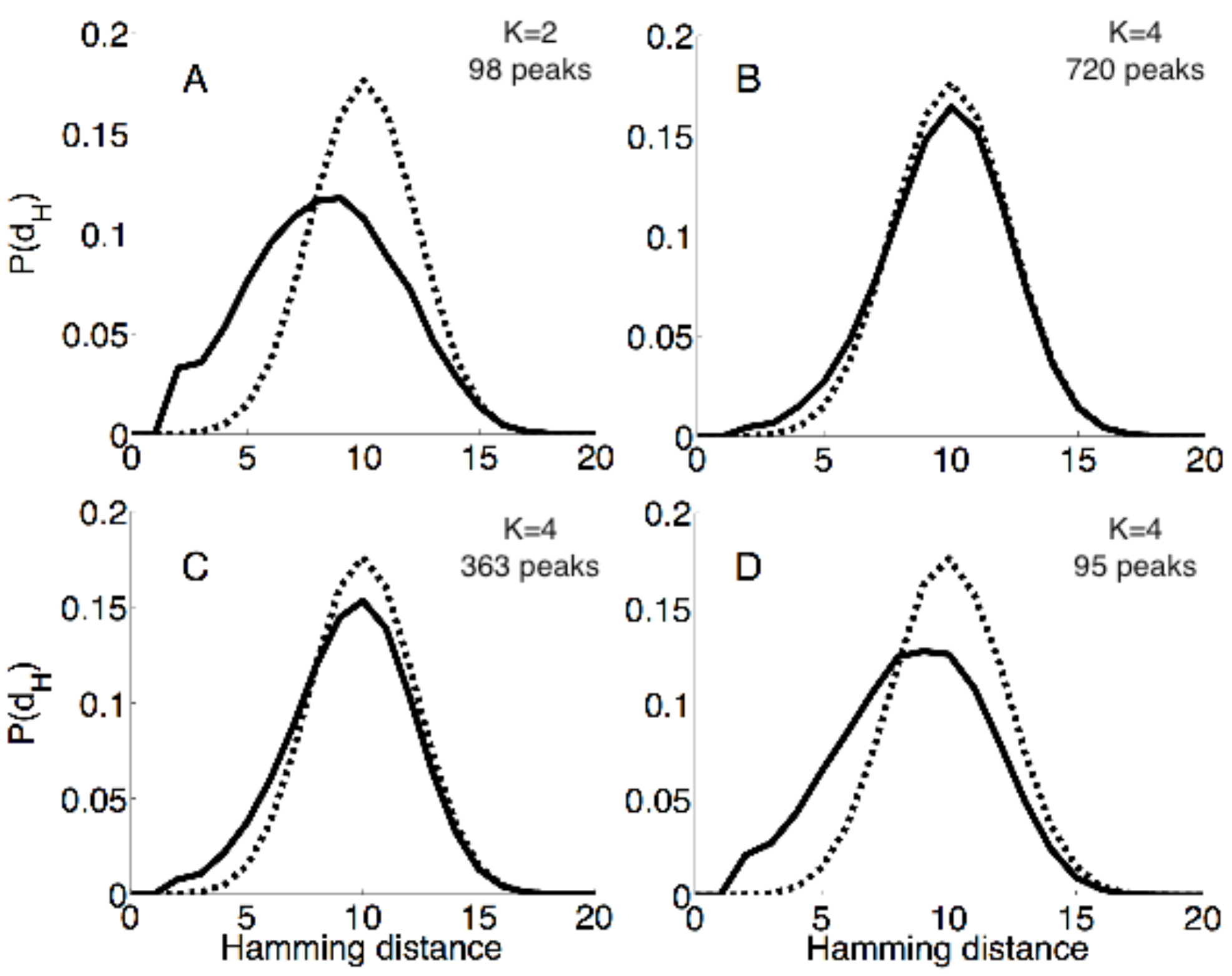}
\caption{Distributions of pairwise Hamming distances $d_H$ between all peaks (solid) and between random control genotypes (dashed) in the NK landscape. The distributions shown are the averages of 50 different landscapes with genomes of length $N$ = 20. (A) $K$ = 2 landscapes containing an average of 98 peaks. (B) $K$ = 4 landscapes containing an average of 720 peaks. (C) $K$ = 4 landscapes including only an average of 363 peaks with a fitness above a threshold: $W\ge\Theta = 0.60$. (D) $K$ = 4 landscapes including only an average of 95 peaks with a fitness above a threshold of $\Theta= 0.66$. As the samples include only the highest peaks, the pairwise distributions of $K$ = 4 landscapes begin to resemble that of the $K$ = 2 landscapes, suggesting that the highest peaks do cluster in genotype space, whereas the distribution of lower peaks is less biased. The control assigns genotypes to peaks by randomly sampling from all genotypes without replacement. Adapted from~\cite{Ostmanetal2010}.}
\label{fig:2}       
\end{figure}

\begin{figure}[t]
\centering
\includegraphics[scale=.65]{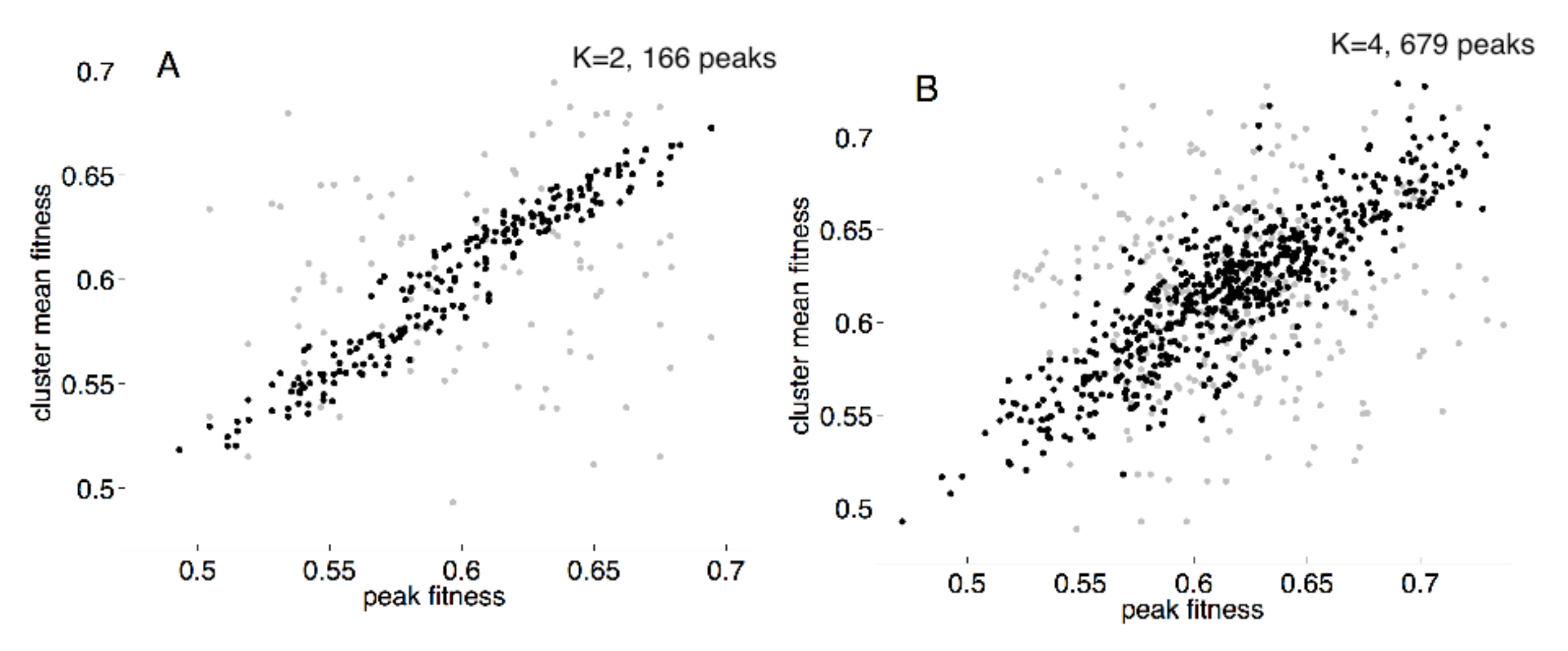}
\caption{Mean fitness of clusters of peaks as a function of peak fitness. The data is obtained comparing the fitness of each peaks with the average fitness of all other peaks within a Hamming distance of $d=2$. (A) Landscape of $K = 2$ with 166 peaks (black dots). All landscapes show a strong correlation between cluster mean fitness and peak fitness, while the same analysis of assigning random genotypes without replacement to the peaks (but keeping the fitness) shows no such correlation (gray dots). The random data are from ten samplings. (B) One landscape of $K = 4$ with 679 peaks (black dots), and random genotypes (gray dots) obtained by sampling four times. Adapted from~\cite{Ostmanetal2010}.}
\label{fig:3}       
\end{figure}

The question of whether populations would be able to utilize the neutral ridges in a Holey Landscape~\cite{Gavrilets1997} can be investigated using a simple computational model. We begin with a homogenous population at the same genotype having a fitness of 1. The object is to find a genotype two mutations away either through a valley or a larger number of mutations away via a neutral ridge (Fig.~6). Each mutation takes the offspring to either one of two neighboring genotypes, such that they can reach the target genotype through two mutations where the first has fitness 0.7, or through four neutral mutations with genotypes of fitness 1. Results for a population size of 100 are shown in Fig.~7, and clearly shows that if the mutation-supply rate is not prohibitively low (as in the strong-selection weak-mutation regime where mutations go to fixation individually) then populations can easily endure deleterious mutations necessary to cross valleys, and will do so rather than a utilizing neutral ridges requiring a larger number of neutral mutations. 

\begin{figure}[t]
\centering
\includegraphics[scale=.5]{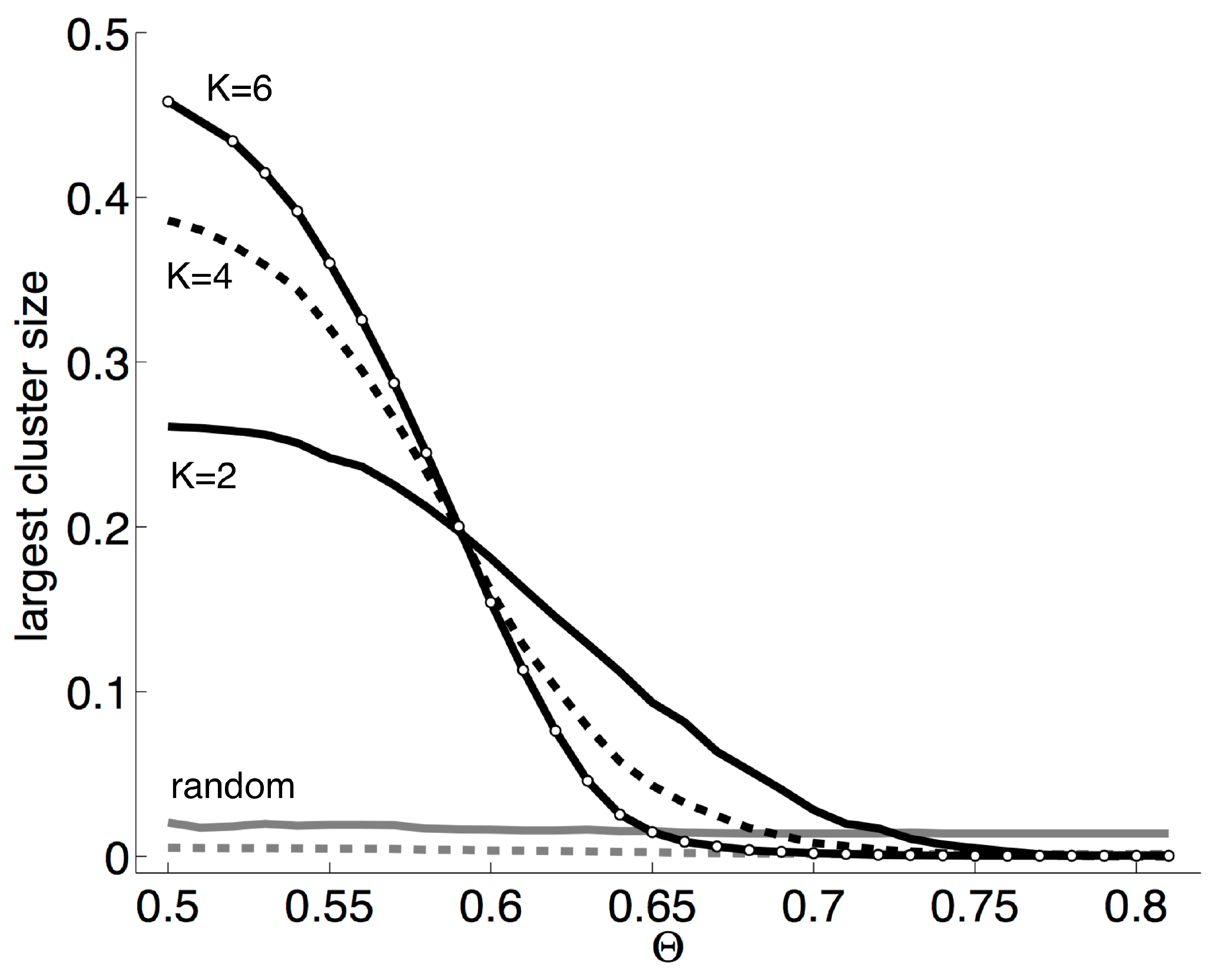}
\caption{Size of the largest network cluster in the landscape averaged over 50 landscapes for $K$ = 2, 4, and 6 as a function of fitness threshold, $\Theta$. Clusters consist of networks of peaks no further than Hamming distance $d$ = 2 from each other. Only peaks with fitness above $\Theta$ are included in each sample. The more rugged the landscapes are, the more abrupt the transition is from small network clusters to one cluster dominating the landscape. Random genotypes for $K$ = 2 (solid gray line) and $K$ = 4 (dashed gray line) show no increase in cluster size as $\Theta$ is lowered, indicating that peaks cluster in a non-random. Adapted from~\cite{Ostmanetal2010}.}
\label{fig:4}       
\end{figure}

\begin{figure}[t]
\centering
\includegraphics[scale=.39]{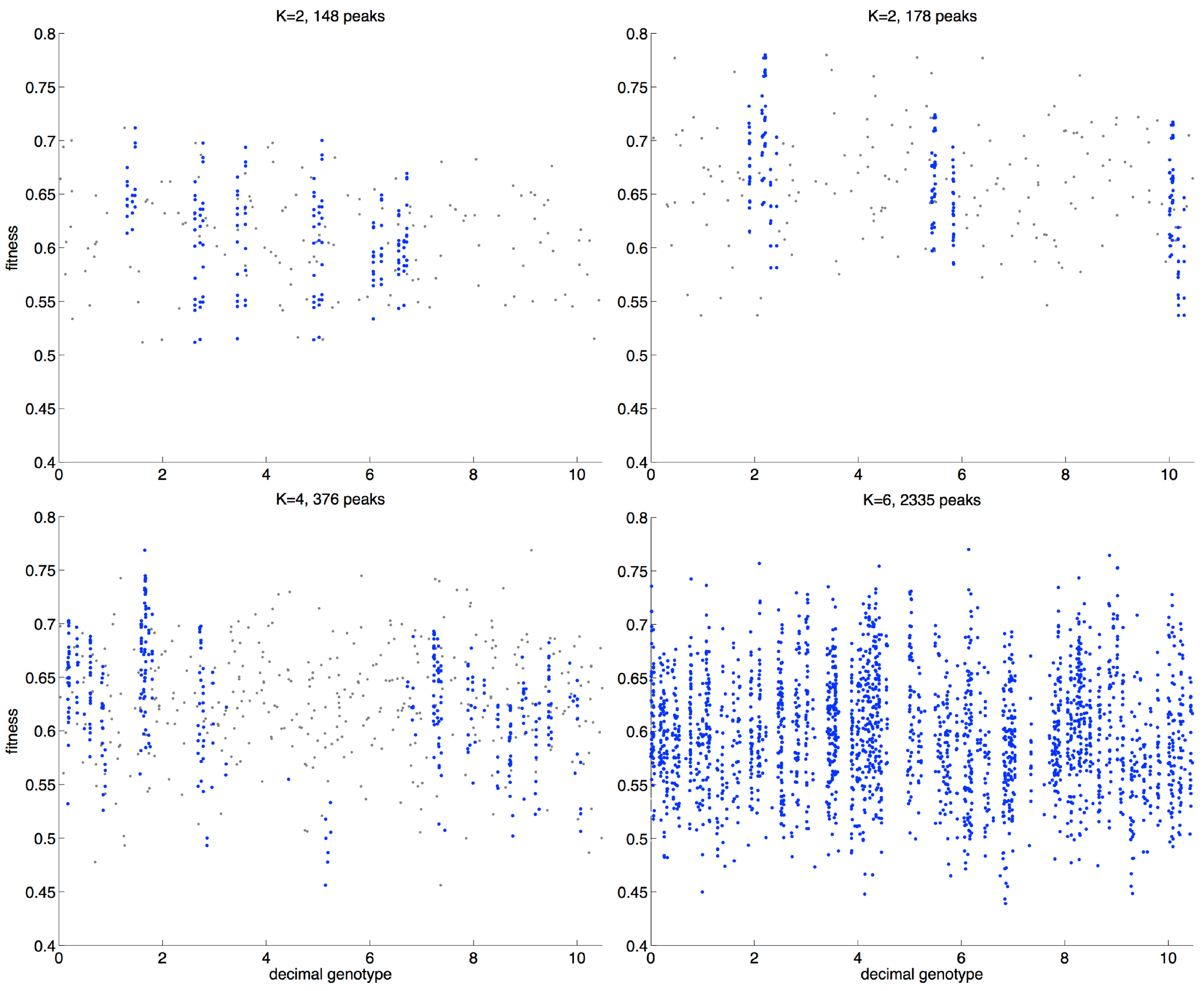}
\caption{Basins of attraction in the NK landscape. The bit-string genome of an NK organism can be translated into a decimal number, and peaks are here plotted with fitness as a function of this decimal genotype. Four different landscapes for $K=2$ (top row) and $K=4$ and 6 (bottom row) show that peaks cluster and huge voids exist with no peaks. Because neighboring genotypes can have very different decimal genotypes, not all adjacent peaks appear close to each other in decimal space. A single mutation at one locus can add a very high number to the decimal genotype, and so it is not immediately clear which peaks are close to each other in actual genotype space. Random reassignment of genotypes to peaks results in a random distribution in decimal genotype space (gray points in $K=2$ and 4).
Other kinds of fitness landscapes can be visualized in a similar fashion, even when the number of ``alleles" higher than two.
}
\label{fig:5}       
\end{figure}

\begin{figure}[t]
\centering
\includegraphics[scale=.74]{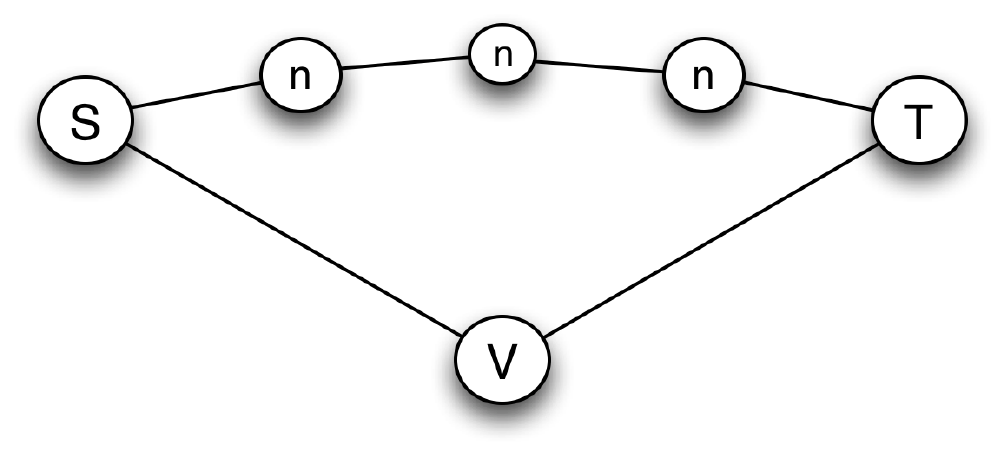}
\caption{Model for computing the probability that a population will traverse a valley rather than taking the path of a neutral ridge. The population starts at genotype S with fitness 1 and can either take the path via genotypes n--which all have fitness 1--or go through the valley via genotype V (with fitness 0.7). The simulation is stopped when the first individual reaches the target genotype T (we record which path the population took to get there).}
\label{fig:6}       
\end{figure}
\begin{figure}[!ht]
\centering
\includegraphics[scale=.35]{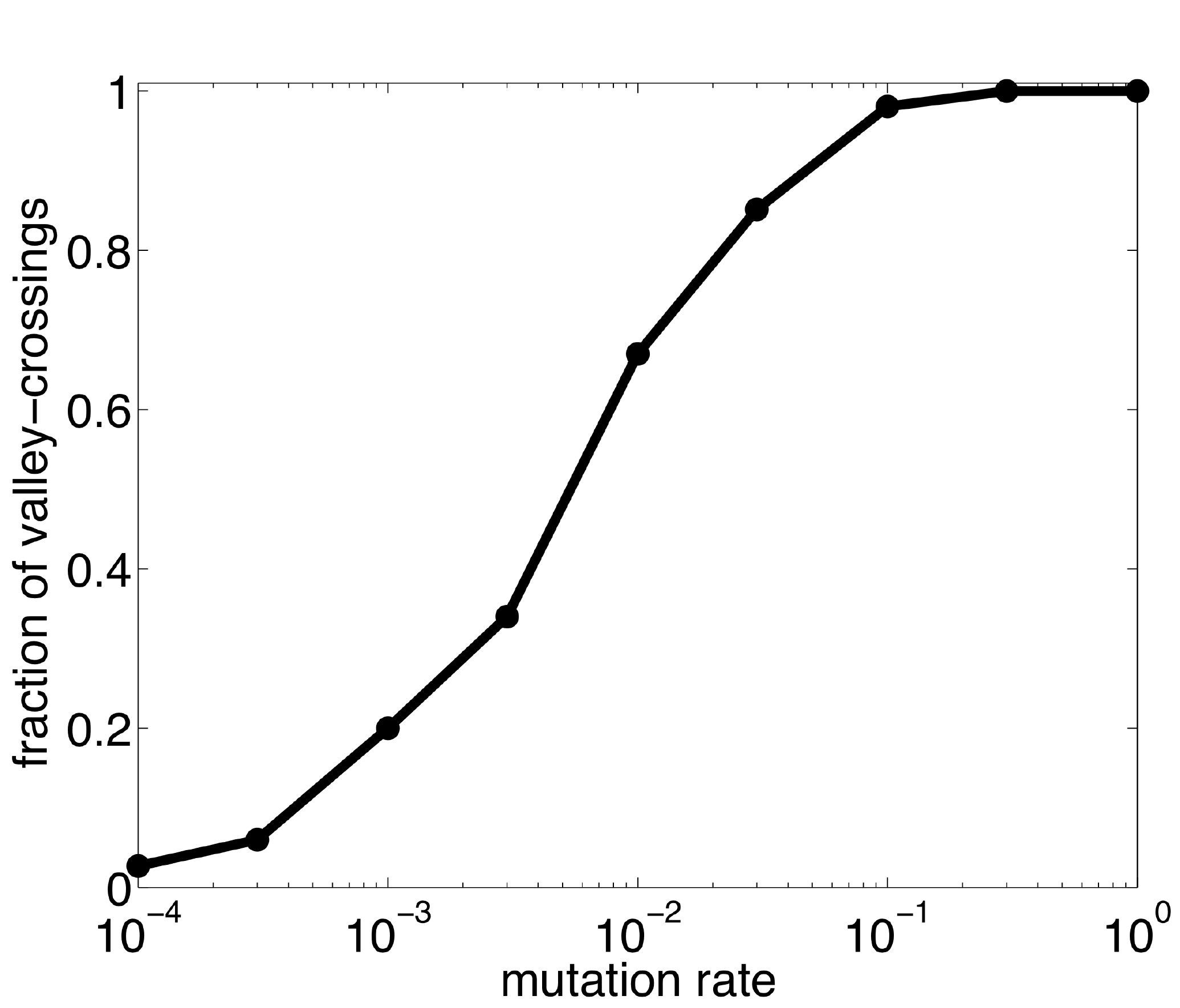}
\caption{Fraction of 100 simulations where the population took the path through a valley instead of a neutral ridge. When the mutation rate is not too low, valley-crossings are frequent. A population of 100 individuals is started with a genotype of fitness 1, and can evolve to the target genotype by taking a neutral path of length 4 or by going through a valley with fitness 0.7. The neutral ridge genotypes all have fitness 1, and traversing the valley requires only two mutations.}
\label{fig:7}       
\end{figure}

\section{Landscape structure and prediction}

Predicting evolution is a notoriously difficult task~\cite{Bush2001}, and the success of such an endeavor depends crucially on the parameters of the evolutionary process (e.g., mutation rate and population size for a non-recombining population) as well as the structure of the fitness landscape. If the mutation supply rate is very low and selection is strong, predicting evolution is comparably easier because the only stochastic component of the process lies in calculating the probability that a beneficial mutation (the one closest to the dominating type) achieves a sufficiently large clone size so that the fixation of that mutation becomes deterministic~\cite{GerrishLenski1998,Orr2000,JohnsonBarton2002}. However, recent evidence suggests that this limit is rarely if ever achieved in nature~\cite{Blountetal2012} and that instead many beneficial, neutral, and deleterious mutations exist within realistic populations at the same time.  Furthermore, selection on the deleterious mutations appears to be weak enough that such mutations can persist in large populations for hundreds or even thousands of generations~\cite{Weissmanetal2009} so that they can serve as stepping stones for further adaptive process. In this limit, analytical methods fail to provide closed form solutions for the probability of fixation, in part because valley-crossing now becomes an important component in predicting evolutionary paths. It is precisely in the limit where landscape structure becomes a crucial element in predicting evolution.
	The results described above suggest not only that high fitness peaks are more likely to be found near peaks of similar fitness, but they also imply that peaks of lower fitness permeate genetic space, so that it is possible to traverse the entire genetic space via paths that connect peaks separated only by a single deleterious mutation. While we do not suggest that such paths are more likely than those that involve possible neutral and beneficial mutations, the observation implies that the areas of fitness with high-fitness peaks are accessible from anywhere in the landscape. The picture painted by Gavrilets is in a sense complementary, where large neutral networks of low fitness provide access to the large neutral networks of high fitness peaks.  But such a picture is neither necessary to understand adaptive progress (because in the limit of strong mutation and weak selection deleterious mutations can persist for a very long time in the population), nor is such a picture likely, because evolutionary paths consisting only of neutral and beneficial mutations are vastly under-represented (compared to those that include valley-crossings) when peaks are clustered. Thus, while the possibility of valley crossings makes evolution muchawa more difficult to predict, prediction is considerably helped if the local structure of the landscape is such that high peaks are near other high peaks, and that peak networks percolate genetic space (albeit only at low fitness). 

\section{Future directions}

While the concept and in particular the metaphor of the fitness landscape now has a long and distinguished history, we still have a lot to learn about the global and local structure of realistic landscapes with epistasis and pleiotropy. This is certainly true for computational models in which the fitness landscape is not explicitly designed, but it is particularly evident for empirical fitness landscapes, which have only recently begun to be investigated in the simplest cases. Because fitness landscape structure remains largely unknown, the predictive power that knowledge of that structure could bring has not yet been explored. 
	Here we have proposed that together with population size and mutation rate, fitness landscape structure is a key factor in enabling prediction of evolutionary outcomes. Despite the natural stochasticity of the evolutionary process, to a first approximation the trajectories that evolving populations take can be predicted with some degree of accuracy. The success of adaptation and the likelihood of speciation can be approximated given information about these three key evolutionary parameters. Because of this prospect, various methods for visualizing critical properties of fitness landscapes are of great interest. The distribution of genotypes within populations depends not only on the supply of mutations, but also on the local structure of the fitness landscape; flat landscapes result in a wider distribution and thus a higher level of genetic variation than populations located on a high peak with steep slopes. The amount of genetic variation in turn affects the evolvability of the population, which is thus intimately tied to the structure of the fitness landscape.
	
	In order to make predictions about adaptation and speciation using fitness landscapes, several features are of particular interest. As outlined above, the spatial distribution of peaks in genotype space will affect evolutionary trajectories, so species-specific knowledge of the local structure must be obtained. While for some organisms it is evident that there are peaks in the local genotype or phenotype neighborhood, it is unclear if this is a general feature of all organisms. It is possible that some or most organisms find themselves at a peak that is prohibitively far from other peaks. In such cases, if the landscape changes deprecating the occupied peak, the environmental change could lead to extinction before the population can respond adaptively. The extent of peak clustering in the local neighborhood thus anticipates the evolutionary outcome. 
Given that speciation can be driven by adaptation to specialized niches represented by distinct peaks in fitness landscapes~\cite{Schluter2001}, it follows that the ruggedness of the landscape and the number of peaks and their proximity to each other influences the likelihood of speciation. From this we can hypothesize a correlation between landscape ruggedness and the rate of speciation. We expect that the more rugged the landscape is, the higher the rate of speciation will be. It may be that too much ruggedness diminishes this effect when peaks are too numerous and close to each other, in which case valleys occupied by hybrids may not prohibit interbreeding, and thus lead to a breakdown of reproductive isolation in sexual species. In this case there would exist an optimum degree of ruggedness that maximizes the rate of speciation. If ruggedness could be estimated, it could be compared between clades that have different rates of speciation. Our expectation would then be that the rate of speciation is an increasing function of ruggedness up to a point where peaks become too numerous, leading to a breakdown in reproductive isolation. However, we currently have no way to estimate local fitness landscape ruggedness without genotyping and measuring fitness on a massive scale.

	We investigated here the question of whether populations would be able to utilize the neutral ridges in a Holey Landscape~\cite{Gavrilets1997} using a computational model (Fig.~6), and show that even a large reduction in fitness (up to 30\%) is tolerated if the mutation rate is not very low. An important variable here is the population size: If the population size is small, the mutation-supply rate may be so small that neutral paths cannot reliably be found. If the population size is large, then even very small fitness differences can be detected by selection, and thus even fewer paths are going to be neutral. It is possible that the available neutral paths in the limit of very high-dimensional genotype space are too few and far between so that bridging peaks along neutral ridges becomes much less likely than valley-crossing. Further investigation into the effect of the length of the neutral pathway is also warranted.
	
It is now clear~\cite{Weissmanetal2009,Ostmanetal2012} that deleterious mutations can segregate and exist in populations for extended periods of time. We also know from experiments with microbes that populations experience beneficial mutations over long periods of time. Indeed, in over 50,000 generations of evolution in the Long Term Evolution Experiment (LTEE)~\cite{Barricketal2009}, populations of {\it E.~coli} continue to adapt with no sign of reaching a fitness plateau~\cite{Wielgossetal2013}. It Is not clear whether this evidence is consistent with the Holey Landscape model of fitness landscapes, in which beneficial mutations are abundant when the population is not at the highest fitness. If beneficial mutations were that easy to find, why are beneficial mutants still being found after over 50,000 generations of evolution? The evidence from the LTEE suggests, instead, that beneficial mutations are rare but continue to be available to lead to evolutionary progress.

Empirical fitness landscapes may turn out to have less in common with computational landscapes than we would have hoped, and it is therefore critical that empirical landscapes are investigated in greater detail. The common approach is to measure fitness of many genotypes a couple of mutations away from the peak, but this is unlikely to yield information about the ruggedness of the landscape. We therefore suggest that it would be valuable to measure the fitness of genotypes that are significantly removed from the wild-type (up to, say, 10 mutations away). This is impossible to do exhaustively because of the very large set of mutants, but selecting a subset to probe the fitness landscape in this manner could provide information about the fitness landscape pertinent to predicting evolutionary dynamics.

	Finally, we acknowledge that real biological landscapes are not static entities as assumed in this work, but can change in both time and space. Evolutionary dynamics in non-static landscapes may be quite different from that of static landscapes~\cite{HaydenWagner2012,MustonenLaessig2009}. For example, pathogenic microbes experience changing fitness landscapes as the host continuously attempts to fight off the infection, and all organisms likewise may at times find certain traits to change from being advantageous to being sub-optimal because of environmental change. In these cases adaptation may be reduced to simply climbing the closest peak, as the previously optimal genotype has shifted, and the population finds itself in a position of low fitness. However, it is not known how frequently such change occurs, or how much it actually affects the fitness landscape. It may be that extreme cases of qualitative changes where peaks shift in genotype space are rare, and that most environmental change results in smaller deprecations of peaks, which has a less dramatic effect on evolution. It may be that many populations find themselves in effectively static landscapes for most of their evolutionary history. We therefore contend that knowledge of evolution in static fitness landscapes will still be relevant for our understanding of evolutionary dynamics.

\section*{Acknowledgement}
We would like to thank Anurag Pakanati and Charles Ofria for extensive discussions on the structure of fitness landscape. This work was was supported in part by the National Science FoundationÕs Frontier in Integrative Biological Research Grant No. FIBR-0527023 and NSFÕs BEACON Center for the Study of Evolution in Action, under Contract No. DBI-0939454. We wish to acknowledge the support of the Michigan State University High Performance Computing Center and the Institute for Cyber Enabled Research.
%
\bibliography{refsforChris}
\bibliographystyle{spphys}
\end{document}